\documentclass[%
 reprint,
 amsmath,amssymb,
 aps,
]{revtex4-1}

\usepackage{graphicx}
\usepackage{dcolumn}
\usepackage{bm}
\usepackage{float} 

\usepackage{subfiles}

\newcommand*{\affaddr}[1]{#1} 
\newcommand*{\affmark}[1][*]{\textsuperscript{#1}}

\newcommand{\be}{\begin{equation}}
\newcommand{\ee}{\end{equation}}
\newcommand{\bea}{\begin{eqnarray}}
\newcommand{\eea}{\end{eqnarray}}
\newcommand{\beq}{\begin{equation}}
\newcommand{\eeq}{\end{equation}}
\newcommand{\beqn}{\begin{eqnarray}}
\newcommand{\eeqn}{\end{eqnarray}}

\begin{document}

\preprint{APS/123-QED}

\title{Probing axion-like particles with $\gamma \gamma$ final states from vector boson fusion processes at the LHC} 

\author{
Andr\'es Fl\'orez\affmark[2], Alfredo Gurrola\affmark[1], Will Johns\affmark[1], Paul Sheldon\affmark[1], Elijah Sheridan\affmark[1], Kuver Sinha\affmark[3], Brandon Soubasis\affmark[1]\\
\affaddr{\affmark[1] Department of Physics and Astronomy, Vanderbilt University, Nashville, TN, 37235, USA}\\
\affaddr{\affmark[2] Physics Department, Universidad de los Andes, Bogot\'a, Colombia}\\
\affaddr{\affmark[3] Department of Physics and Astronomy, University of Oklahoma, Norman, OK 73019, USA}\\
}

\date{\today}

\begin{abstract}

We perform a feasibility study to search for axion-like particles (ALPs) using vector boson fusion (VBF) processes at the LHC. We work in an effective field theory framework with cutoff scale $\Lambda$ and  ALP mass $m_{a}$, and assume that ALPs couple to photons with strength $\propto 1/\Lambda$. Assuming proton-proton collisions at $\sqrt{s} = 13$ TeV, we present the total VBF ALP production cross sections, ALP decay widths and lifetimes, and relevant kinematic distributions as a function of $m_{a}$ and $\Lambda$. We consider the $a\to\gamma\gamma$ decay mode to show that the requirement of an energetic diphoton pair combined with two forward jets with large dijet mass and pseudorapidity separation can significantly reduce the Standard Model backgrounds, leading to a $5\sigma$ discovery reach for $10 \text{ MeV} \lesssim m_{a} \lesssim 1$ TeV with $\Lambda \lesssim 2$ TeV, assuming an integrated luminosity of 3000 fb$^{-1}$. In particular, this extends the LHC sensitivity to a previously unstudied region of the ALP parameter space.
\end{abstract}

\pacs{Valid PACS appear here}
\maketitle

\section{\label{sec:level1}Introduction}

A major current focus of searches beyond the Standard Model (SM) is the QCD axion \cite{Weinberg:1977ma, Wilczek:1977pj, Peccei:1977hh} and more general pseudo-scalar axion-like particles (ALPs) $a$, which are ubiquitous in string theory \cite{Arvanitaki:2009fg, Cicoli:2012sz}. These hypothetical particles are being probed by a wide array of methods that exploit the ALP-photon coupling~\cite{Sikivie:1983ip,Sikivie:1985yu,Anastassopoulos:2017ftl,Irastorza:2013dav,Spector:2019ooq}.  We refer to Ref.~\cite{Graham:2015ouw} for a review of these topics. While astrophysical searches (for example using neutron stars  \cite{Fortin:2021sst, Harris:2020qim, Lloyd:2020vzs, Fortin:2018ehg, Fortin:2018aom, Buschmann:2019pfp}) provide a potentially rich avenue for constraining ALPs, lab-based searches are particularly important given the control over both production and detection. Such searches include beam dump \cite{Dobrich:2015jyk} and fixed target experiments including FASER \cite{Feng:2018pew}, LDMX \cite{Berlin:2018bsc,Akesson:2018vlm}, NA62 \cite{Volpe:2019nzt}, SeaQuest \cite{Berlin:2018pwi}, and SHiP \cite{Alekhin:2015byh}; newer proposals, such as PASSAT~\cite{Bonivento:2019sri, Dev:2021ofc}, which are hybrids of a beam dump and a  helioscope; and even more recently proposed reactor neutrino-based ideas \cite{Dent:2019ueq}, \cite{Kelly:2020dda}.

Laboratory-based searches for ALPs coupling to photons fall into several distinct categories depending on the regions of parameter space they are sensitive to. Light-shining-through-wall experiments \cite{Spector:2019ooq}, which rely on ALP-photon conversions, probe smaller ALP masses $m_a \lesssim 10^{-3}$ eV; beam dumps that rely on ALP decay typically probe larger masses $ \sim \mathcal{O}(1$ MeV - $1$ GeV$)$ ; while hybrid proposals like PASSAT  probe an intermediate regime $m_a \lesssim 100$ eV. 

High energy colliders are sensitive to a large swathe of the ALP mass and ALP-photon coupling parameter space. Theoretical studies of ALPs  at the LHC and future colliders arising from on-shell decays $h \rightarrow aa$,  $h \rightarrow Za$ and $Z \rightarrow \gamma a$ have been performed by several authors \cite{Jaeckel:2015jla, Brivio:2017ije, Bauer:2018uxu, Alves:2016koo}. Constraints from LEP arise from associated production of ALPs via $e^+e^- \rightarrow \gamma a \rightarrow 3 \gamma$ and $e^+e^- \rightarrow Z \rightarrow \gamma a \rightarrow 3 \gamma$. On the other hand, exotic decays of the Higgs and the $Z$ form the basis of many LHC searches via $pp \rightarrow h \rightarrow Za \rightarrow Z\gamma \gamma$ and  $pp \rightarrow h \rightarrow aa \rightarrow 4 \gamma$.

\begin{figure}[]
 \begin{center} 
 \includegraphics[width=0.4\textwidth]{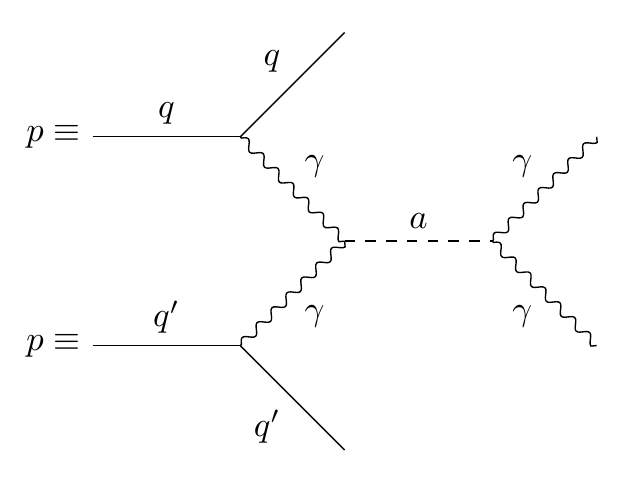}
 \end{center}
 \caption{ALP VBF production diagram}
 \label{fig:feynDY}
 \end{figure}

The purpose of this paper is to perform a careful investigation of ALPs at the LHC arising from photon fusion processes utilizing the vector boson fusion (VBF) topology, and assuming that ALPs couple to SM photons. The relevant Feynman diagram is shown in Fig.~\ref{fig:feynDY}. An early study in this direction was performed by the authors of \cite{Jaeckel:2012yz} using LHC data from 2011 and 2012. In the mass window $100$ GeV $< \, m_a \, < 160$ GeV, ATLAS VBF Higgs searches \cite{vbfatlas1, Aad:2012tfa} were used to establish upper limits on the allowed signal cross section in each $m_{a}$ bin. For higher masses, the constraints were directly obtained by comparing the observed number of events in the diphoton mass spectrum over the expected background distribution, while for lower masses down to $m_a \sim 50$ GeV, ATLAS measurements of photon pair production were used \cite{Aad:2012tba}. 

In our work, we will perform an updated study of ALPs using the VBF topology, down to ALP masses at the MeV scale below which they decay outside the detector. The VBF topology has been proposed as an effective tool for a variety of beyond-SM searches, such as dark matter~\cite{Florez:2019tqr, DMmodels2, CMSVBFDM}, supersymmetry~\cite{VBF1,VBFSlepton,VBFStop,VBFSbottom,VBF2,Sirunyan:2019zfq,ConnectingPPandCosmology}, $Z'$~\cite{VBFZprime}, heavy neutrinos~\cite{VBFHN} and heavy spin-2 resonances \cite{Florez:2018ojp}. As we will show, it is particularly effective for probing ALPs. Our results are summarized in  Fig.~\ref{fig:compare}, where we see that VBF enables sensitivity to a regime of parameter space that is not covered by any other experiment.

\section{Samples and simulation}

In the case of the axion 
signal samples, the model files were generated using the FeynRules package~\cite{Alloul:2013bka} and obtained from Ref.~\cite{axionUFO}. The interactions between $a$ and SM particles are described by a 5-dimensional operator in the Lagrangian, where the kinetic term $\mathcal{L} \supset e^{2}C_{\gamma\gamma}\frac{a}{\Lambda}F_{\mu\nu}\tilde{F}^{\mu\nu} + \frac{2e^{2}}{s_{w}c_{w}}C_{\gamma Z}\frac{a}{\Lambda}F_{\mu\nu}\tilde{Z}^{\mu\nu} + \frac{e^{2}}{s^{2}_{w}c^{2}_{w}}C_{ZZ}\frac{a}{\Lambda}Z_{\mu\nu}\tilde{Z}^{\mu\nu}$ represents the $a$ interactions with 
the photon ($\gamma$) and $Z$ boson. 
The Wilson coefficients $C_{\gamma\gamma}$, $C_{\gamma Z}$, and $C_{ZZ}$ govern the $a\to\gamma\gamma$, $a\to\gamma Z$, and $a\to ZZ$ decays, respectively. In the above Lagrangian, $s_{w}$ and $c_{w}$ are the sine and cosine of the weak mixing angle, $F_{\mu\nu}$ and $Z_{\mu\nu}$ the energy-momentum tensors of $\gamma$ and $Z$, and $\Lambda$ the symmetry breaking scale. We produced several signal samples considering various values of $\mathcal{A}\equiv \frac{C_{ij}}{\Lambda^2}$. For the purpose of the studies shown in this paper, we set the value of the coefficients $C_{ij}$ to unity, but scenarios with different values can be derived by appropriately re-scaling the production cross-sections. 

Simulated events from proton-proton ($pp$) collisions at $\sqrt{s}=13$ TeV were generated for signal and background using MadGraph5\_aMC (v2.6.5) 
~\cite{MADGRAPH}. Hadronization was performed with PYTHIA (v8.2.05) \cite{Sjostrand:2014zea}. Detector effects were included through Delphes (v3.4.1) \cite{deFavereau:2013fsa}, using the CMS input card. 

\begin{figure}
    \centering
    \includegraphics[width=.5\textwidth, height=.25\textheight]{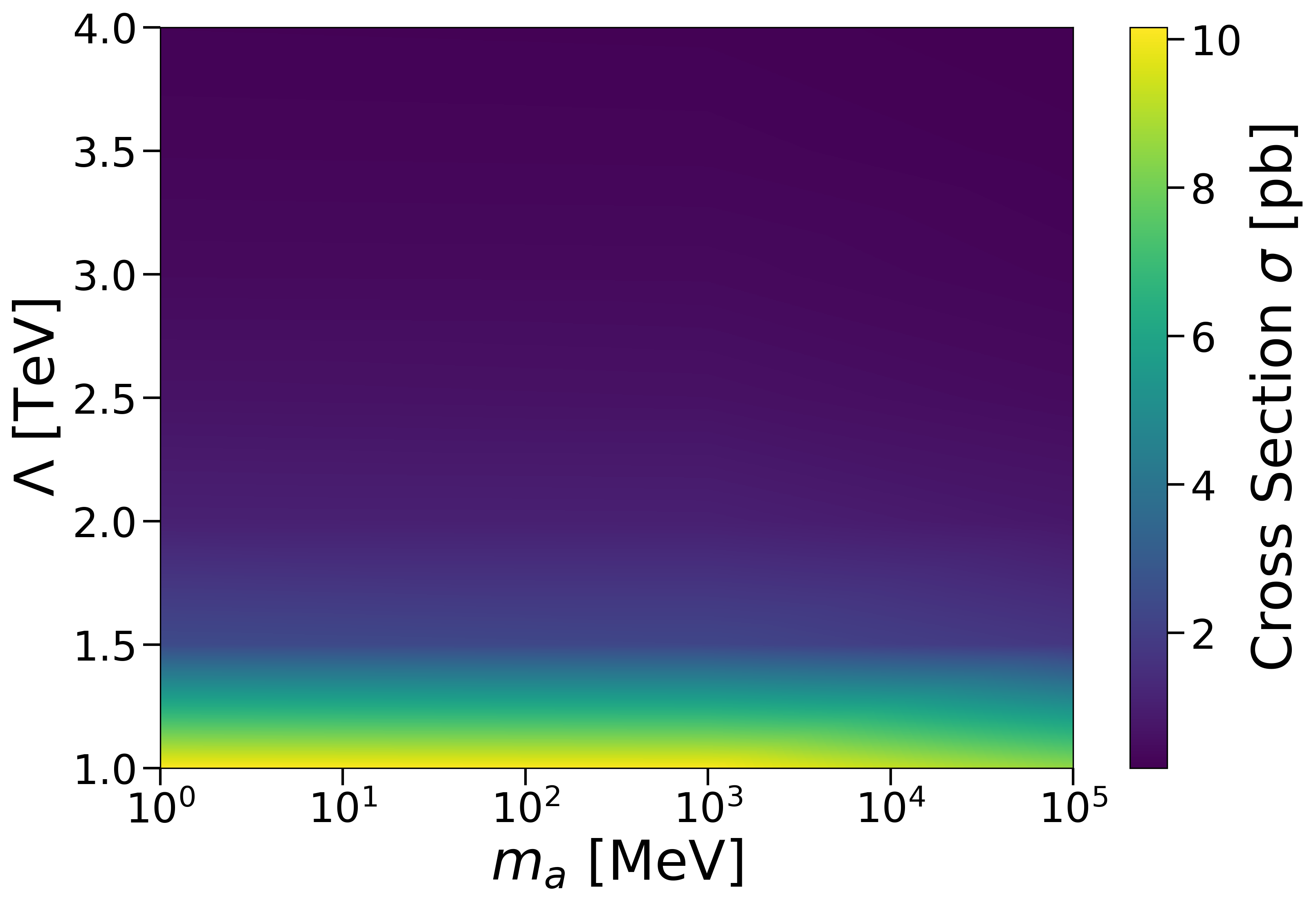}
    \caption{The VBF $a jj$ cross section as a function of $m_{a}$ and $\Lambda$.}
    \label{VBFcrosssection}
\end{figure}

The signal samples were produced for a variety of axion masses, ranging from 1 MeV to several TeV. 
The value of $\Lambda$ was varied between 1000 GeV to 4000 GeV, for every ALP mass point generated. Pure electroweak production of a ALP and two additional jets (i.e. $pp\rightarrow a jj$ with suppressed QCD coupling $\alpha_{QCD}^{0}$) was considered. At MadGraph level, jets were required to have a minimum $p_{T} > 20$ GeV, $|\eta| < 5$, a pseudorapidity gap of $|\Delta\eta_{jj}| > 2.4$, and reconstructed dijet mass of $m_{jj} > 120$ GeV. The parton level $\Delta\eta_{jj}$ and $m_{jj}$ requirements reduce the contributions from s-channel gluon-gluon fusion ($gg\to a$) and associated ALP production diagrams (e.g., $q\bar{q}\to Z^{*} \to Z a \to j j a$), which can result in a similar final state, in order to optimize the VBF $ajj$ statistics in our samples. Figure~\ref{VBFcrosssection} shows the $ajj$ production cross section, with the parton level requirements described above, as a function of $m_{a}$ for varying values of $\Lambda$. 
Photons were required to have transverse momentum greater than 10 GeV and located in the central region of the ATLAS and CMS detectors ($|\eta (\gamma)| < 2.5$). Photon pairs were also required to be separated in $\eta$-$\phi$ space by requiring $\Delta R^{\gamma \gamma} = \sqrt{(\Delta \phi^{\gamma\gamma})^2+(\Delta \eta^{\gamma\gamma})^2} >0.4$.

\begin{figure}[]
 \begin{center} 
 \includegraphics[width=0.45\textwidth]{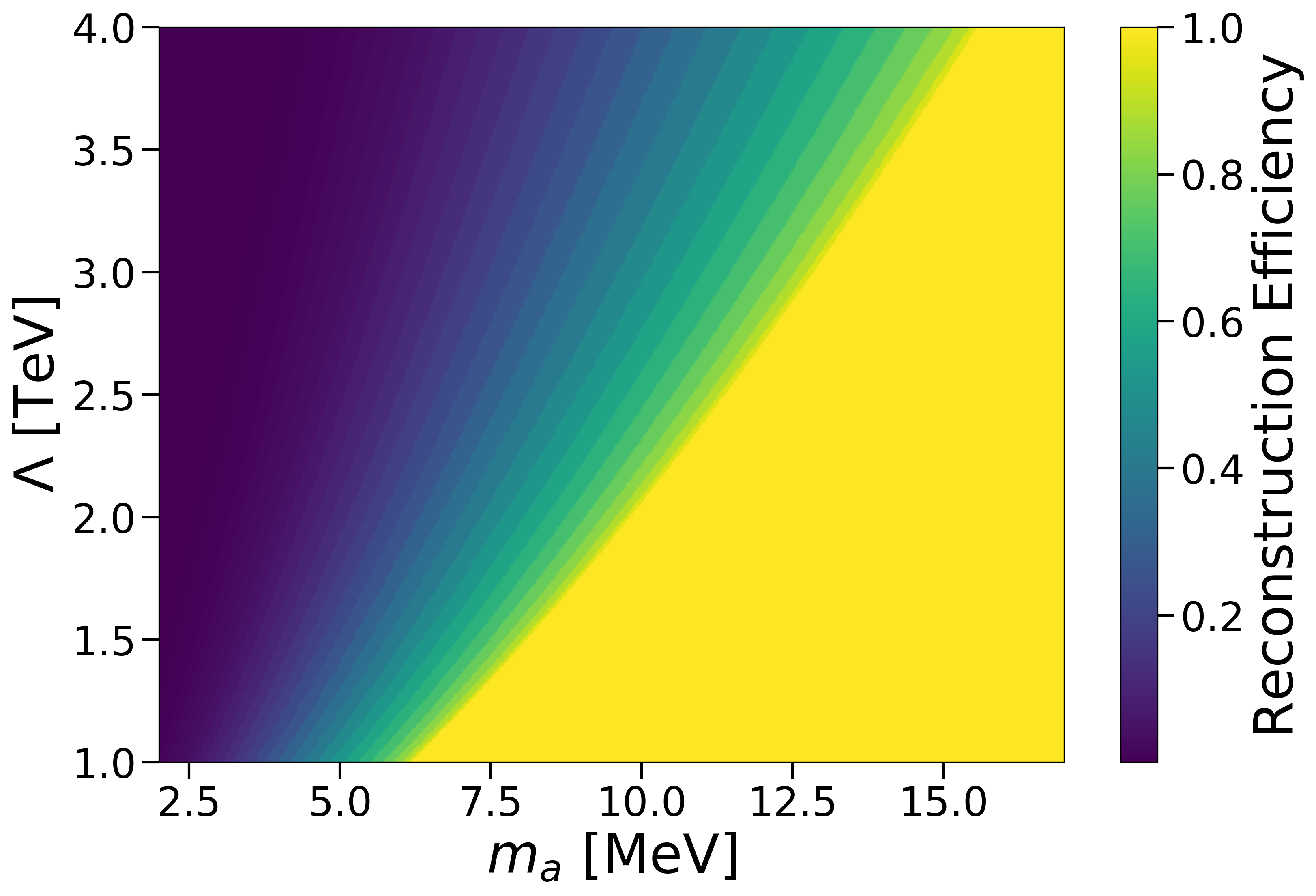}
 \end{center}
 \caption{The fraction of events which decay inside the detector and leave a signature in the CMS ECAL, as a function of $m_{a}$ and $\Lambda$.}
 \label{fig:decaylength}
 \end{figure}

We note that the resonant ALP production cross-section via VBF is given by $\sigma_{VBF} \propto \frac{m_{a}^{2}}{\Lambda^{2}}$, and is thus suppressed for relatively small ALP masses with respect to the symmetry breaking scales considered in these studies. For this reason, non-resonant ALP production dominates the cross-section in a large part of the $m_{a}$ phase space considered, a property observed and exploited by the authors of \cite{Gavela:2019}. Similarly, the ALP decay width $\Gamma$ is suppressed by $m_{a}$ over the new physics scale $\Lambda$, and thus ALPs with small $m_{a}$ can be long-lived and decay outside of the detector. To determine the range in $m_{a}$ at which the long lifetime becomes important, we compute the ALP decay length perpendicular to the proton-proton beam axis, which has the form $L_{a,\perp} = \frac{\sqrt{\gamma_{a}^{2}-1}}{\Lambda}\sin\theta$. In this equation, $\theta$ is the scattering angle relative to the beam axis and $\gamma_{a}$ is the relativistic boost factor. This quantity is calculated per simulated signal event by utilizing the ALP pseudorapidity distribution and conservatively assuming the ALP moves at the speed of light. Since our focus is on the $a\to\gamma\gamma$ decay channel, events with $L_{a,\perp}$ values corresponding to an ALP decay beyond the CMS electromagnetic calorimeter (ECAL) cannot be used, so we neglect regions of the ALP parameter space where this happens to a non-trivial extent (see Figs. \ref{fig:results_lum150} and \ref{fig:results_lum3000}). Figure~\ref{fig:decaylength} shows the fraction of events which decay inside the detector and leave a signature in the CMS ECAL, as a function of $m_{a}$ and $\Lambda$. For $\Lambda = 1$ TeV ($4$ TeV), a large fraction of the events are lost when $m_{a} < 5$ MeV ($m_{a} < 15$ MeV) .

The dominant sources of SM background are production of photon pairs with associated jets, referred to as $\gamma\gamma$+jets. In the proposed search region (defined in Section III), the associated jets are mainly from initial state radiation (i.e. $pp \to \gamma \gamma jj$, $\alpha_{QCD}^{2}$) or SM VBF processes (i.e. $pp \to \gamma \gamma jj$, $\alpha_{QCD}^{0}$). Therefore, the background samples are split into two categories: $(i)$ non-VBF $\gamma\gamma$+jets events with up to four associated jets, inclusive in the electroweak coupling ($\alpha_{EWK}$) and $\alpha_{QCD}$; and $(ii)$ pure electroweak $\gamma\gamma jj$. The production of $\gamma$+jets and multijet events with jets misidentified as photons have been checked to provide a negligible contribution to the proposed search region due to the effectiveness of the VBF selection criteria. 

The MLM algorithm \cite{MLM} was used for jet matching and jet merging. 
The xqcut and qcut variables of the MLM algorithm, related with the minimal distance between partons and the energy spread of the clustered jets, were set to 30 and 45 as result of an optimization process requiring the continuity of the differential jet rate as a function of jet multiplicity. 

 \section{Event selection criteria}

We focus on a final state with exactly two well identified photons and two jets consistent with the characteristics of the photon-photon fusion process. Stringent requirements are placed on the $p_{T}$ of photons, and on the kinematic properties of the VBF dijet system in order to suppress SM backgrounds. 

\begin{figure}[]
    \centering
    \includegraphics[width=0.45\textwidth]{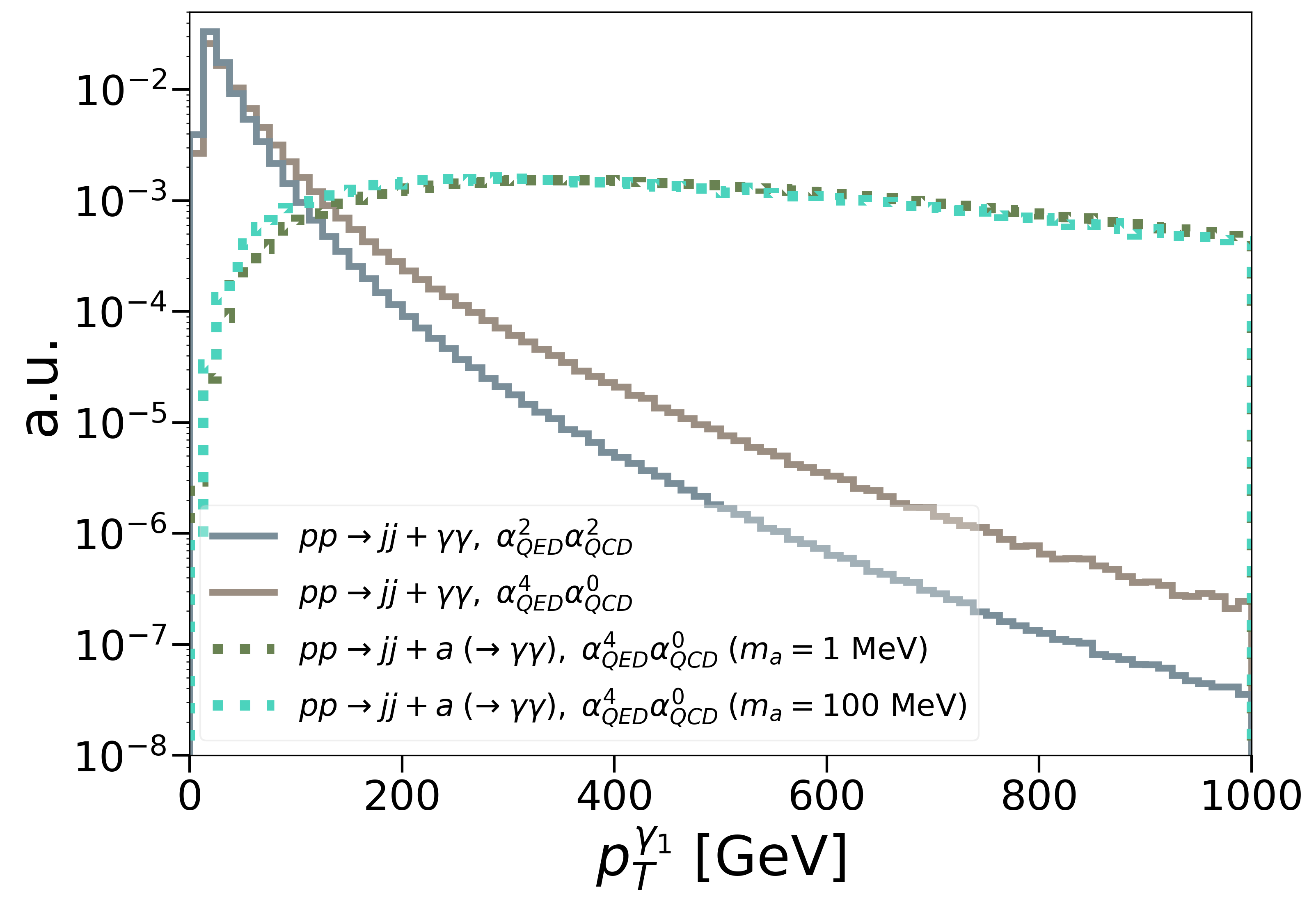}
    \caption{The leading photon transverse momentum mass distribution (normalized to unity) for the total SM backgrounds and $m_a = 1$ MeV, $m_a = 100$ MeV signal benchmark points.}
    \label{fig:pta1_distr}
\end{figure}

To study the important differences between signal and background processes, we select events with at least two $\gamma$ candidates with $|\eta^\gamma| < 2.5$ and $p^{\gamma} > 10$ GeV, and present various kinematic distributions. The $\gamma$ with the highest $p_{T}$ is referred to as the leading $\gamma$. Figure~\ref{fig:pta1_distr} shows the leading photon transverse momentum distribution, $p_{T}^{\gamma_{1}}$, for two signal benchmark samples and the main associated backgrounds, normalized to the area under the curve (unity). Note the signal protrudes around $p_{T}^{\gamma_{1}} > 200$ GeV, but the exact cut value of $p_{T}^{\gamma_{1}} > 300$ GeV is determined through an optimization process aimed at maximizing discovery potential. The optimization of all cut values was performed using the statistical figure of merit $N_{S}/\sqrt{N_{S}+N_{B}+(0.25 \times (N_{B} + N_{S}))^{2}}$, where $N_{S}$ and $N_{B}$ represent the expected number of signal and background events, and the term $0.25 \times (N_{B} + N_{S})$ corresponds to the associated systematic uncertainty on the background plus signal prediction, which is a realistic uncertainty based on VBF searches at ATLAS and CMS~\cite{VBF2,CMSVBFDM,Sirunyan:2019zfq}. We note this particular definition of signal significance is only used  for  the  purpose  of  optimizing  the  selections. The final discovery reach is determined with a shape based analysis (described later) using the full diphoton mass or dijet mass spectrum.

\begin{figure}[]
    \centering
    \includegraphics[width=0.45\textwidth]{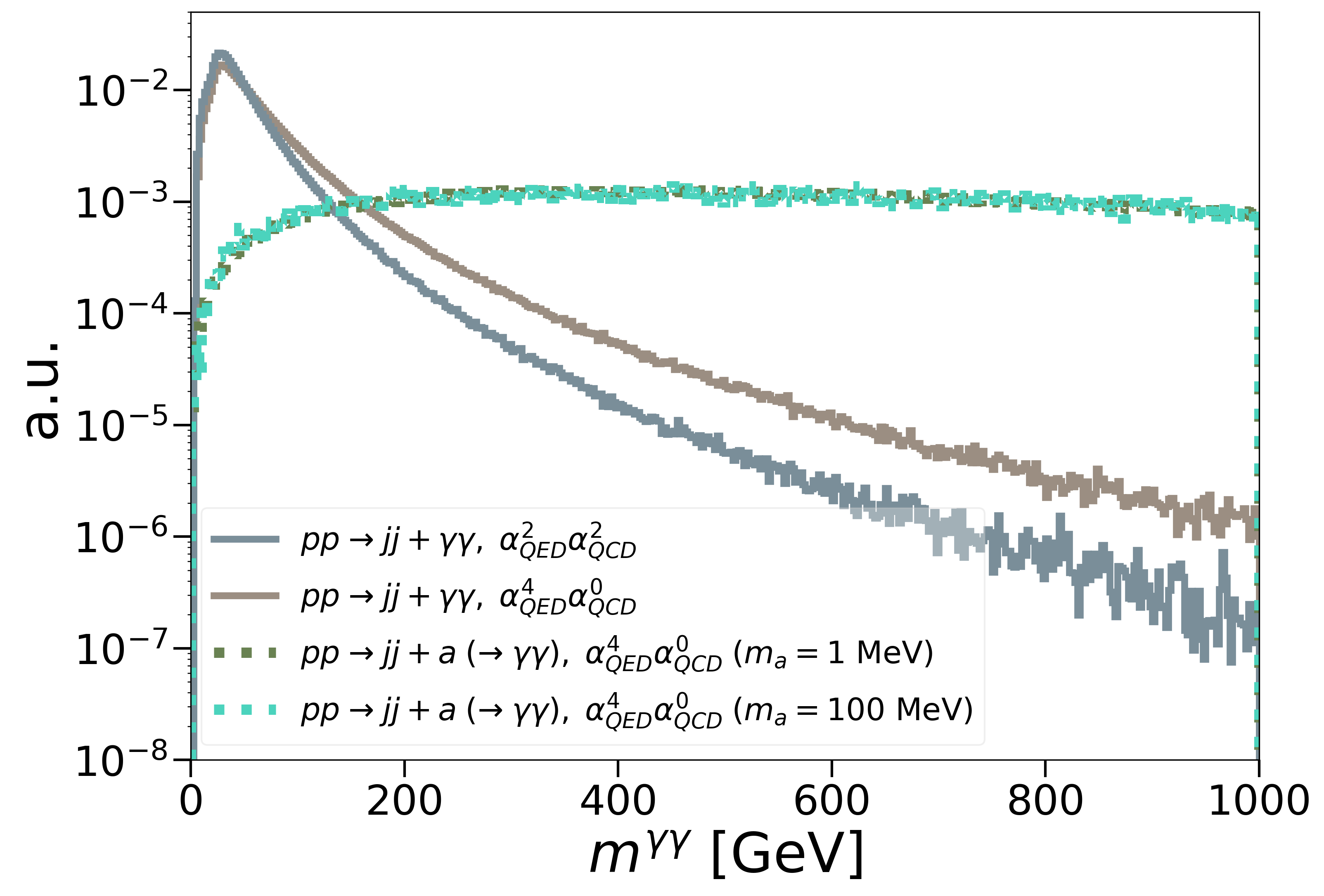}
    \caption{The diphoton mass distribution (normalized to unity) for the total SM backgrounds and $m_a = 1$ MeV, $m_a = 100$ MeV signal benchmark points.}
    \label{fig:maa_distr}
\end{figure}

For low $m_{a}$ values, the relatively large photon $p_{T}$ is a key feature attributed to the kinematically boosted topology facilitated by the VBF process. This kinematic feature provides a nice handle to reconstruct and identify low $m_{a}$ signal events amongst the large SM backgrounds. Figure~\ref{fig:maa_distr} shows the reconstructed mass of the photon pair, $m^{\gamma \gamma}$, normalized to unity, for the SM backgrounds and two signal benchmark points. In the case of non-resonant low mass ALP production, the diphoton mass values scale as $m^{\gamma \gamma} \approx p_{T}^{\gamma_{1}} + p_{T}^{\gamma_{2}}$. Therefore, the high-$p_{T}$ signal photons produce a broad $m^{\gamma \gamma}$ distribution that overtakes the SM backgrounds at several hundred GeV. Since $m^{\gamma \gamma}$ in signal and background events depends on the $p_{T}$ of photons and their angular correlations, we perform a two-dimensional optimization of the $m^{\gamma \gamma}$ and $p^{\gamma_{1}}_{T}$ cut values. Figure~\ref{fig:select_sig_maa_pta1} shows the signal significance 
for $p_{T}^{\gamma_{1}}$ as a function of $m^{\gamma \gamma}$, for a benchmark point with $m_a = 1$ MeV and $\Lambda = 1$ TeV. We select events with $m^{\gamma \gamma} > 500$ GeV. These results were obtained after optimizing the VBF dijet selections (discussed below) 
in order to account for the correlation to the boosted kinematics. 

\begin{figure}[]
    \centering
    \includegraphics[width=0.45\textwidth]{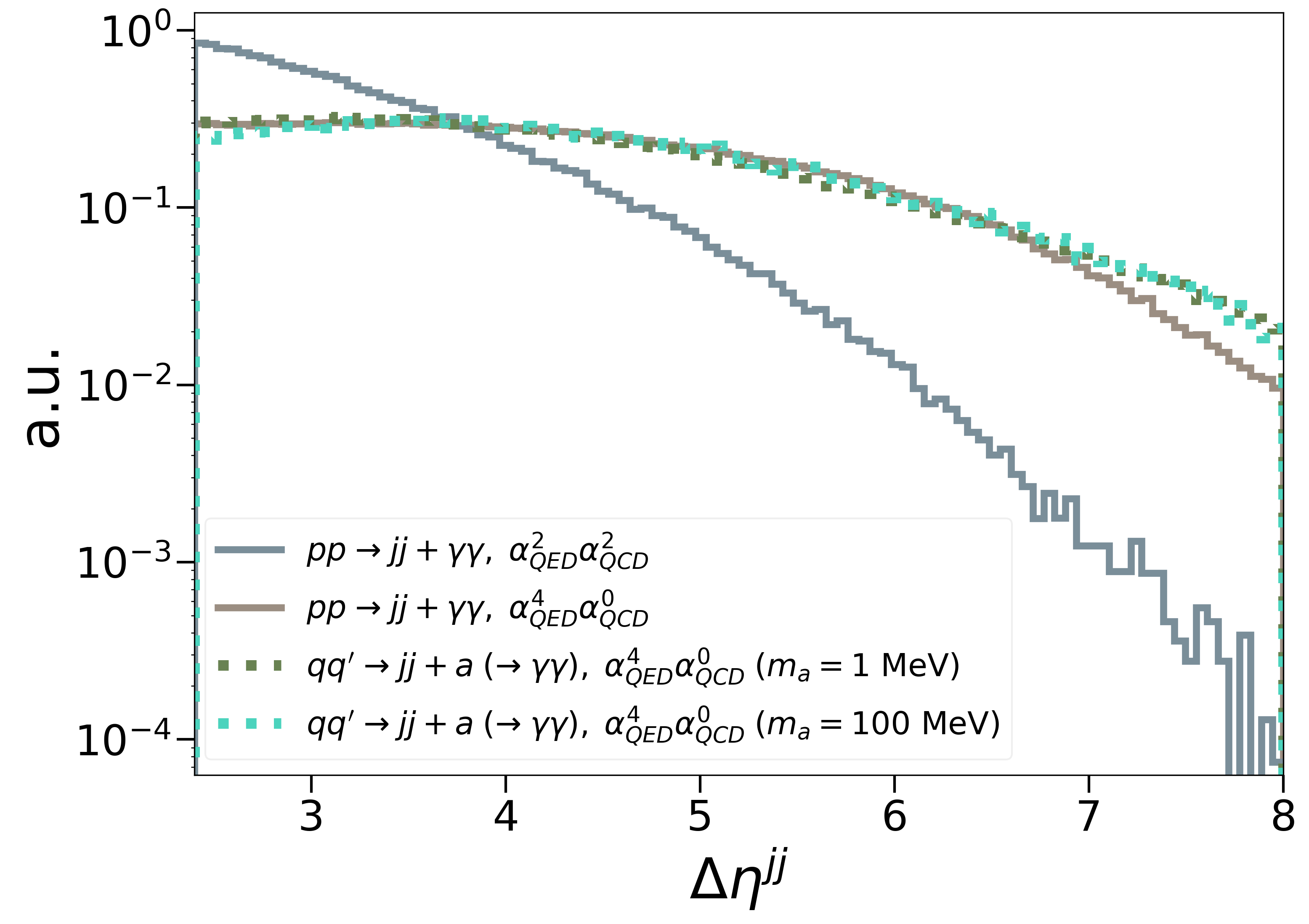}
    \caption{The distribution of the scalar difference in pseudorapidity between jets (normalized to unity) for the total SM backgrounds and $m_a = 1$ MeV, $m_a = 100$ MeV signal benchmark points.}
    \label{fig:deltaeta_distr}
\end{figure}

\begin{figure}[]
    \centering
    \includegraphics[width=0.45\textwidth]{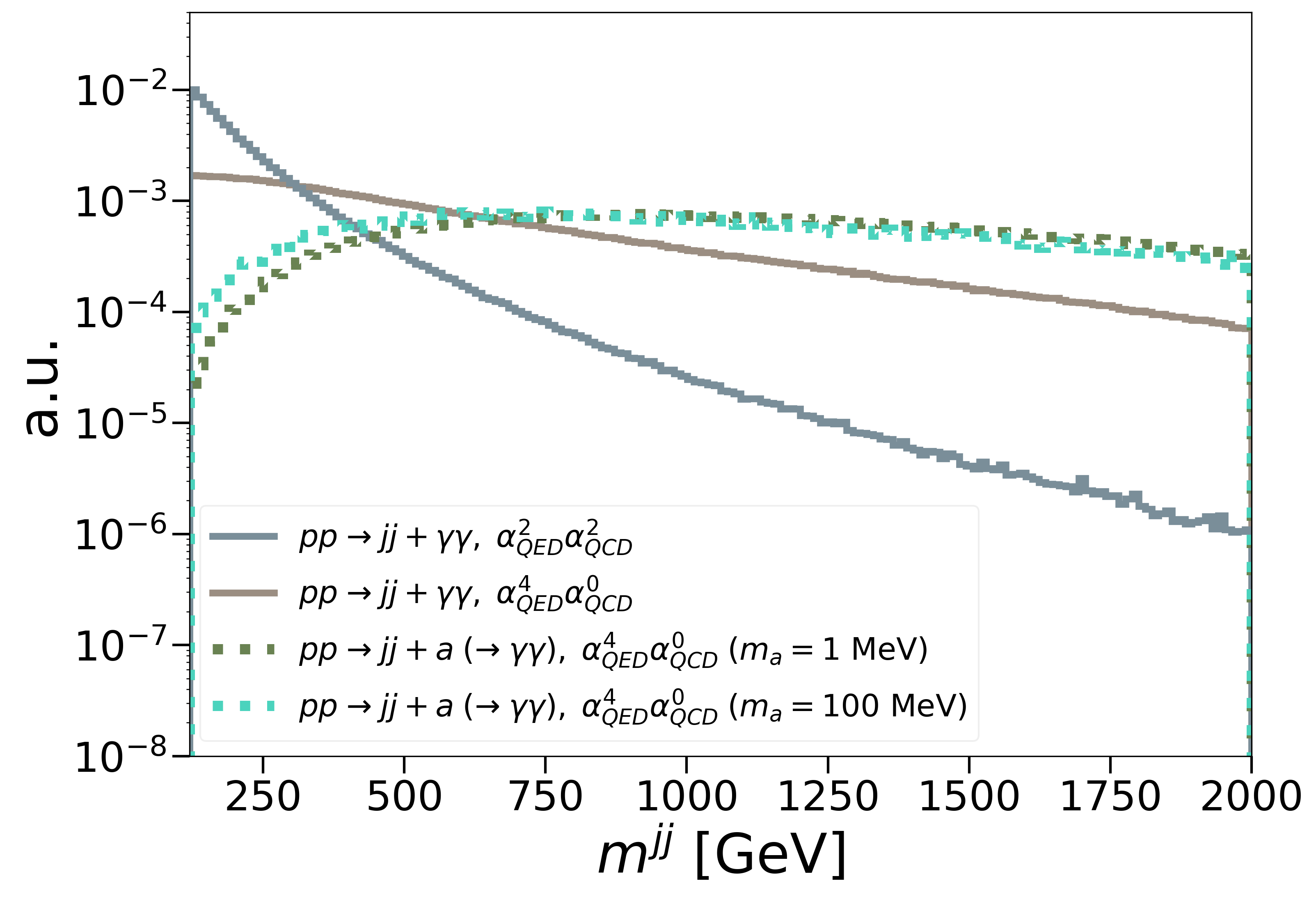}
    \caption{The dijet mass distribution (normalized to unity) for the total SM backgrounds and $m_a = 1$ MeV, $m_a = 100$ MeV signal benchmark points.}
    \label{fig:mjj_distr}
\end{figure}

\begin{figure}
    \centering
    \includegraphics[width=0.45\textwidth]{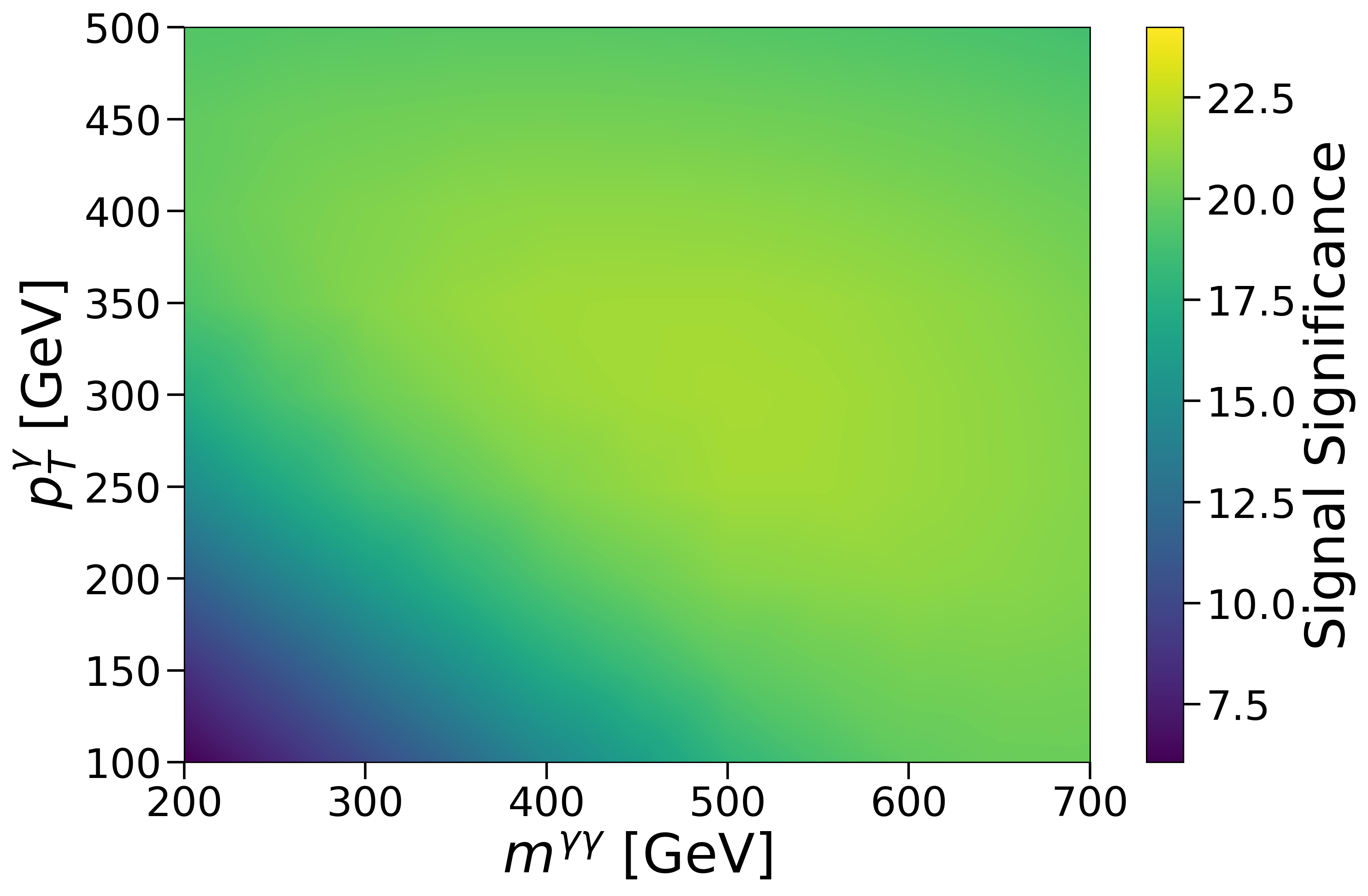}
    \caption{Significance versus selections for the variables $m^{\gamma\gamma}$ and $p_T^{\gamma_1}$, given initial VBF optimized selections $m^{jj} > 1250$ GeV, $|\Delta \eta^{jj}| > 3.6$ ($m_a = 1$ MeV and $\Lambda = 1$ TeV benchmark).}
    \label{fig:select_sig_maa_pta1}
\end{figure}

VBF events are characterized by two forward jets with high $p_{T}$, residing in opposite hemispheres of the detector volume, $\eta_{j_{1}} \times \eta_{j_{2}} <$ 0, containing a large separation in pseudorapidity, $|\Delta \eta^{jj}|$, and large reconstructed dijet mass ($m^{jj}$). For a particle collider such as the LHC, the energy of a jet is very high with respect to the mass of its associated parton, allowing us to approximate the dijet mass as $m^{jj} \approx \sqrt{2p_{T}^{j_{1}}p_{T}^{j_{2}}\text{cosh}(\Delta\eta^{jj})}$. Since the reconstructed $p_{T}$ and $\eta$ values of jets inside the ATLAS and CMS experiments are limited by the performance and geometry of their detectors, the VBF kinematic distributions are studied with a pre-selection of at least two jets with $|\eta| <$ 5.0 and minimum $p_{T}^j >$ 30 GeV. These jets are required to be well separated from photons, by imposing a $\Delta R^{\gamma j} = \sqrt{(\Delta \phi^{\gamma j})^{2} + (\Delta \eta^{\gamma j})^{2}} > 0.4$ requirement. Figure~\ref{fig:deltaeta_distr} shows the $\Delta \eta^{jj}$ distribution for signal and background, normalized to unity, while Fig.~\ref{fig:mjj_distr} shows the corresponding $m^{jj}$ distribution. For events where there are more than two well reconstructed and identified jet candidates, the dijet pair with the larger value of $m^{jj}$ is used in Fig.~\ref{fig:mjj_distr}. The s-channel $\gamma \gamma$ fusion production of signal events results in events with larger $\Delta \eta^{jj}$ separation with respect to background events, and subsequently larger dijet mass spectrum.

\begin{figure}
    \centering
    \includegraphics[width=0.45\textwidth]{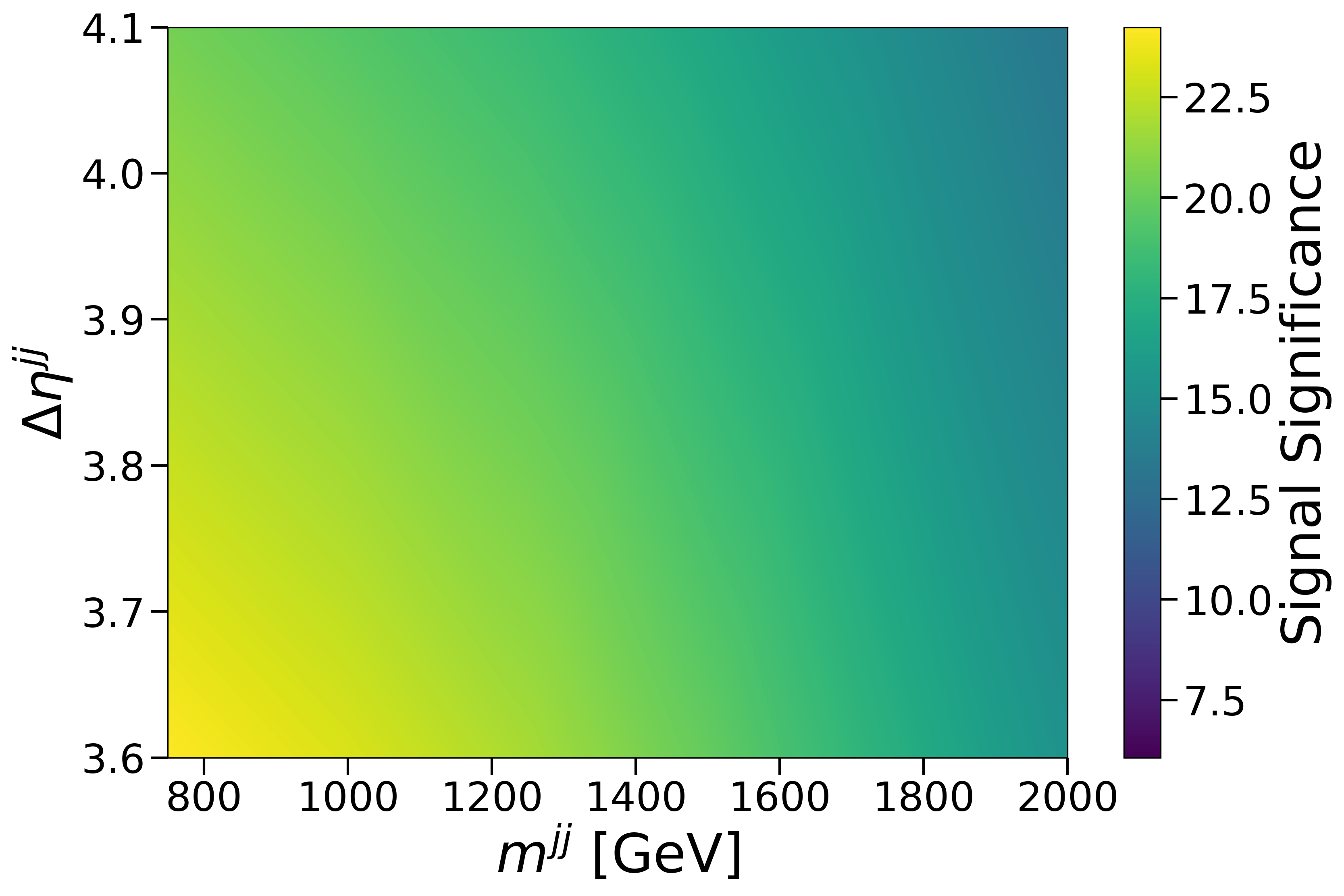}
    \caption{Significance versus selections for the variables $m^{jj}$ and $|\Delta \eta^{jj}|$, given the optimized selections $m^{\gamma\gamma} > 500$ GeV, $p_T^{\gamma_1} > 300$ GeV ($m_a = 1$ MeV and $\Lambda = 1$ TeV benchmark).}
    \label{fig:select_sig_maa_delta_etajj}
\end{figure}

Similar to the optimization of the $p^{\gamma_{1}}$ and $m^{\gamma \gamma}$ requirements, we account for the correlation between $|\Delta \eta^{jj}|$ and $m^{jj}$ by performing a two-dimensional optimization of the $m^{\gamma \gamma}$ and $p^{\gamma_{1}}_{T}$ cut values utilizing the same signal significance definition $N_{S}/\sqrt{N_{S}+N_{B}+(0.25 \times (N_{B} + N_{S}))^{2}}$. To reduce non-VBF signal processes such as gluon-gluon initiated production or associated ALP production such as $Za \rightarrow Z\gamma \gamma \rightarrow jj\gamma \gamma$, we pre-select events with $|\Delta \eta^{jj}| > 3.6$ and $m^{jj} > 750$ GeV. These requirements result in $> 95$\% purity of genuine VBF signal events. Fig.~\ref{fig:select_sig_maa_delta_etajj} shows signal significance as a function of $|\Delta \eta^{jj}|$ and $m^{jj}$.

\begin{table}[]
\begin{center}
\caption {Event selection criteria.}
\label{tab:selection_criteria}
\begin{tabular}{ l  c c}\hline\hline
Criterion & $\gamma_1\gamma_2 j_{f}j_{f}$\\
 \hline
  \multicolumn{3}{ c }{{\bf Central Selections}} \\
   \hline
   $|\eta^{\gamma}|$ & & $< 2.5$ \\
   $p_{T}^{\gamma}$ & & $> 30$ GeV \\
   $p_{T}^{\gamma_{1}}$ & & $> 300$ GeV \\
   $m^{\gamma \gamma}$ & & $> 500$ GeV \\
   \hline
   \multicolumn{3}{ c }{{\bf VBF Selections}} \\
   \hline
   $p_{T}(j)$ & & $>30$ GeV\\
   $|\eta(j)|$ & & $< 5.0$\\
   $\Delta R(\gamma, j)$ & & $> 0.4$\\
   $N(j)$ & & $\geq$ 2\\
   $\eta(j_{1})\cdot \eta(j_{2})$ & & $< 0$\\
   $|\Delta \eta (j_{1}, j_{2})| $ & & $> 3.6$  \\
   $m_{jj}$ & & $> 750.0$ GeV \\
   \hline\hline
 \end{tabular}
\end{center}
\end{table}

\begin{figure}
    \centering
    \includegraphics[width=0.45\textwidth]{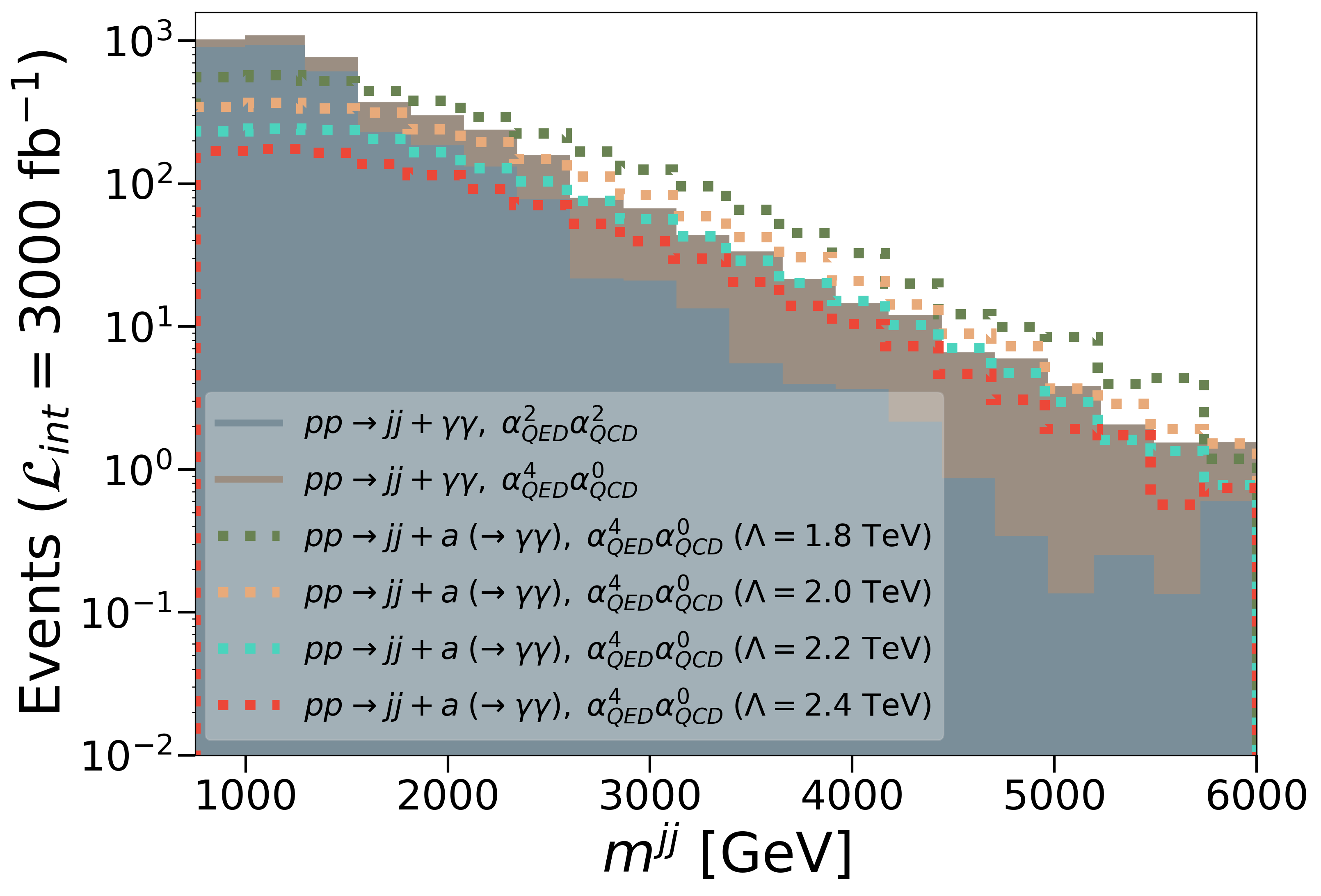}
    \caption{Dijet mass distribution (normalized to cross section, assuming $3000$ fb$^{-1}$ of data) for various signal benchmarks ($m_a = 100$ MeV and $\Lambda$ taking on various values resulting in points near the discovery potential contour in the $m_a$-$\Lambda$ parameter space) after optimized selections imposed on jet and photon kinematic variables.}
    \label{fig:select_stacked_mjj}
\end{figure}

Finally, to completely eliminate other smaller SM backgrounds with top quarks and heavy vector bosons, we impose b-jet and lepton veto requirements. Events are rejected if a jet with $p_{T} > 30$ GeV and $|\eta|< 2.4$ is identified as a bottom quark ($b$). Events are also rejected if they contain isolated electrons or muons with $p_{T} > 10$ GeV and $|\eta|< 2.5$. These requirements are $> 95$\% efficient for VBF ALP signal events. The final optimized event selection criteria is summarized in Table I. Figure~\ref{fig:select_stacked_mjj} shows the expected background and signal yields in bins of $m^{jj}$. Various signal benchmark points are considered and the yields are normalized to cross section times integrated luminosity of $3000$ fb$^{-1}$. The background distributions are stacked/added on top of each other, while the signal distributions are overlaid on the background. 

\section{Results}\label{sec:Results}

To assess the expected experimental sensitivity of this search at the LHC, we followed a profile binned likelihood test statistic approach, using the expected bin-by-bin yields in the reconstructed $m^{\gamma \gamma}$ and $m^{jj}$ distributions. 

Under this approach, the signal significance is defined using the local p-value, understood as the probability of obtaining the same test statistic estimated with the signal plus background hypothesis and from the statistical fluctuation of the background only hypothesis. Then, the signal significance $S$ corresponds to the point at which the integral of a Gaussian distribution between the $S$ and  $\infty$ results in a value equal to the local p-value. The sensitivity was calculated considering the integrated luminosity already collected by ATLAS and CMS experiments during the so called Run-II phase, 150 fb$^{-1}$, and for the 3000 fb$^{-1}$ expected by the end of the LHC era. The estimation of this shape based signal significance was performed using the ROOFit \cite{ROOTFit} toolkit, developed by CERN.

The calculation considers various sources of systematic uncertainties, based upon experimental and theoretical constrains. These uncertainties were incorporated in the test statistic as nuisance parameters. We considered experimental systematic uncertainties on $\gamma$ identification and on reconstruction and identification of jets. For $\gamma$ identification, a conservative 15\% was assumed, following results reported in Ref.~\cite{Khachatryan:2016hje,Aad:2014cka}. The uncertainties between the two photons, and between signal and background process, were considered to be fully correlated. For experimental uncertainties related with the tagging of VBF jets, a 20\% value was included (independent of $m^{jj}$ or $m^{\gamma\gamma}$), following the experimental results from Refs.~\cite{VBF2,CMSVBFDM}. In addition, theoretical uncertainties were included in order to account for the set of parton distribution functions (PDF) used to produce the simulated  signal and background samples. The PDF uncertainty was calculated following the PDF4LHC prescription~\cite{Butterworth:2015oua}, and results in a 5-12\% systematic uncertainty, depending on the process. The effect of the chosen PDF set on the shape of the $m^{jj}$ and $m^{\gamma \gamma}$ distributions is negligible. 

\begin{figure}[]
 \begin{center} 
 \includegraphics[width=0.45\textwidth]{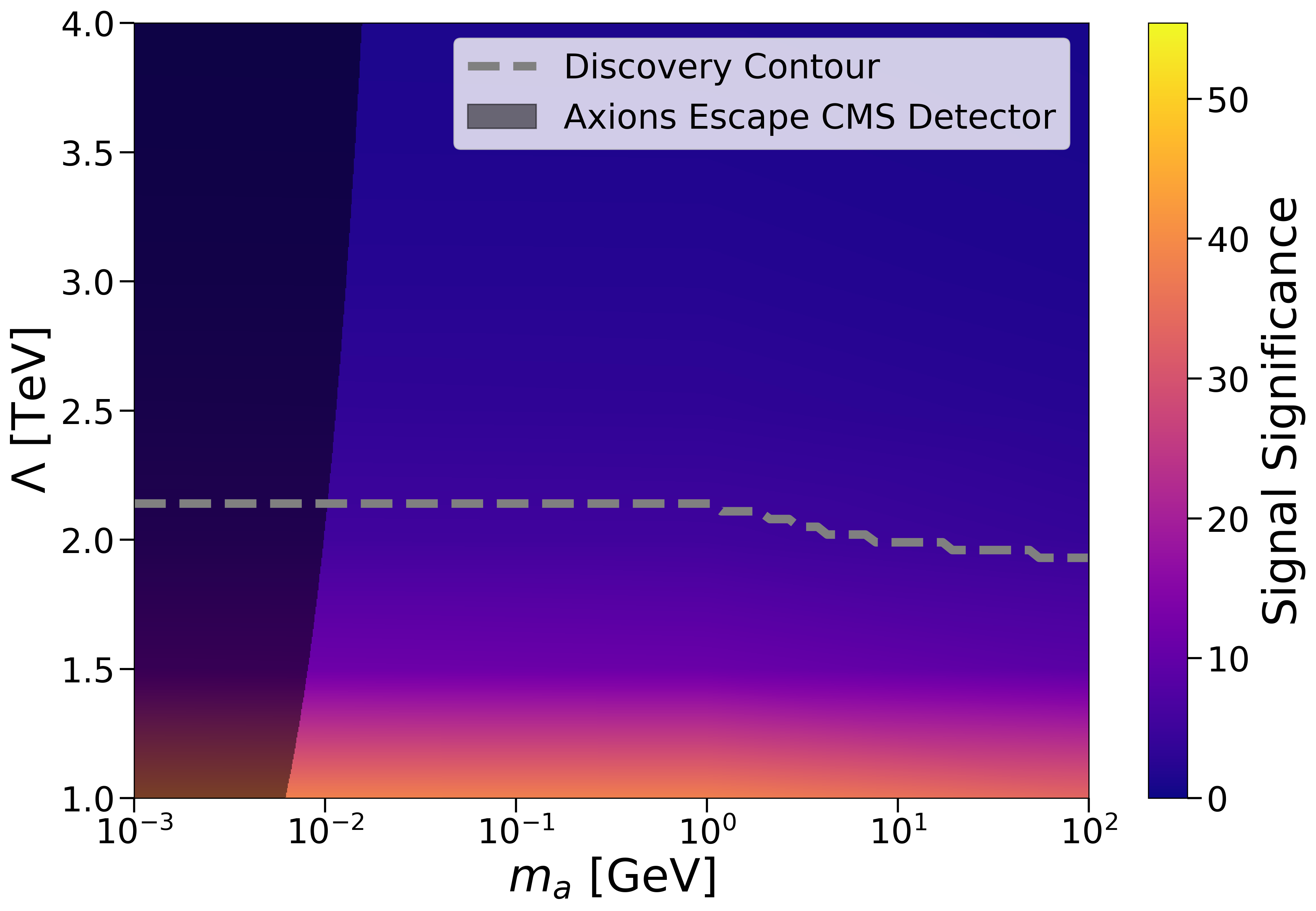}
 \end{center}
 \caption{Expected signal significance for the proposed VBF final state. The results are shown as $m_a$ vs. $\Lambda$ on the $x-y$ plane, and the expected signal significance on the $z-$axis.  The expected signal significance was calculated by interpolating discrete data points as a function of $m_a, \Lambda$, assuming an expected luminosity of 150 fb$^{-1}$. The grey dashed line encloses the region with 5$\sigma$ discovery potential.}
 \label{fig:results_lum150}
 \end{figure}

\begin{figure}[]
 \begin{center} 
 \includegraphics[width=0.45\textwidth]{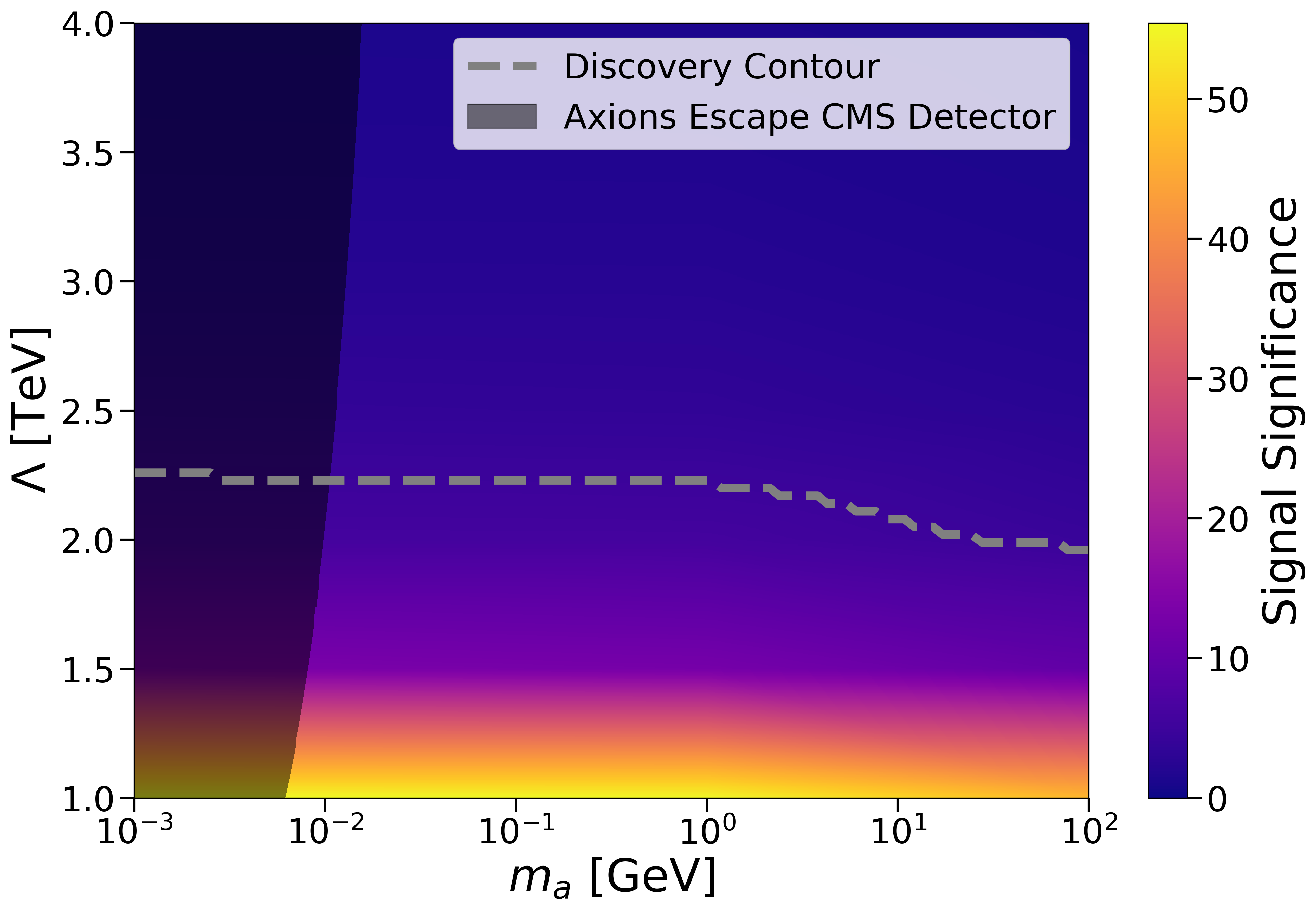}
 \end{center}
 \caption{Expected signal significance for the proposed VBF final state. The results are shown as $m_a$ vs. $\Lambda$ on the $x-y$ plane, and the expected signal significance on the $z-$axis.  The expected signal significance was calculated by interpolating discrete data points as a function of $m_a, \Lambda$, assuming an expected luminosity of 3000 fb$^{-1}$. The grey dashed line encloses the region with 5$\sigma$ discovery potential.}
 \label{fig:results_lum3000}
 \end{figure}

Figures~\ref{fig:results_lum150} and~\ref{fig:results_lum3000} show the results on the expected signal significance for different  $\Lambda$ and $m_{a}$ scenarios, specifically focusing on the lower $m_{a}$ range below $100$ GeV. The dashed line delimits the discovery region.

For the 150 fb$^{-1}$ scenario, it is feasible to probe ALP masses $10 \text{ MeV} \lesssim m_a \lesssim 100$ GeV for $\Lambda \lesssim 1.8\text{-}2.2$, with the latter bound for $\Lambda$ varying with $m_a$. 
The grey band on the plot shows the scenarios in which ALPs decay outside the CMS detector volume, so no detection is possible. 

Similarly, for the 3000 fb$^{-1}$ scenario, the discovery reach includes $10 \text{ MeV} \lesssim m_a \lesssim 100$ GeV for $\Lambda \lesssim 2.0\text{-}2.3$, the $\Lambda$ bound again depending upon $m_a$. The expected discovery reach using the VBF topology includes sensitivity to a regime of the ALP parameter  space  that  is  not covered by any other experiment. This feature is further explained in the following section.

 \section{Discussion}

We have presented a feasibility study for the detection of axion-like particles with strong coupling to photons, $a \rightarrow \gamma \gamma$, produced through VBF processes at the CERN LHC. The expected experimental sensitivity of the search was presented for two different luminosity scenarios, 150 fb$^{-1}$, the current integrated luminosity collected by ATLAS and CMS experiments, and the 3000 fb$^{-1}$ expected by the end of the LHC era. The signal model was developed under an effective field theory approach, considering the symmetry breaking scale, $\Lambda$, and the ALP masses as free parameters.  The expected signal significance for the 150 fb$^{-1}$ scenario allows the ATLAS and CMS experiments to probe ALP masses from 10.0 MeV to 100.0 GeV, for 
values of $\Lambda$ up to 1.8-2.2 TeV, depending on $m_{a}$. For the 3000 fb$^{-1}$ scenario, the discovery reach goes from 10.0 MeV to 100.0 GeV, for 
values of $\Lambda$ up to 2.0-2.2 TeV, depending on $m_{a}$. For $\Lambda$ values below a few TeV, the sensitivity to $m_{a}$ extends to TeV scale values (see Fig. 13, discussed below). 

\begin{figure}
    \centering
    \includegraphics[width=0.45\textwidth]{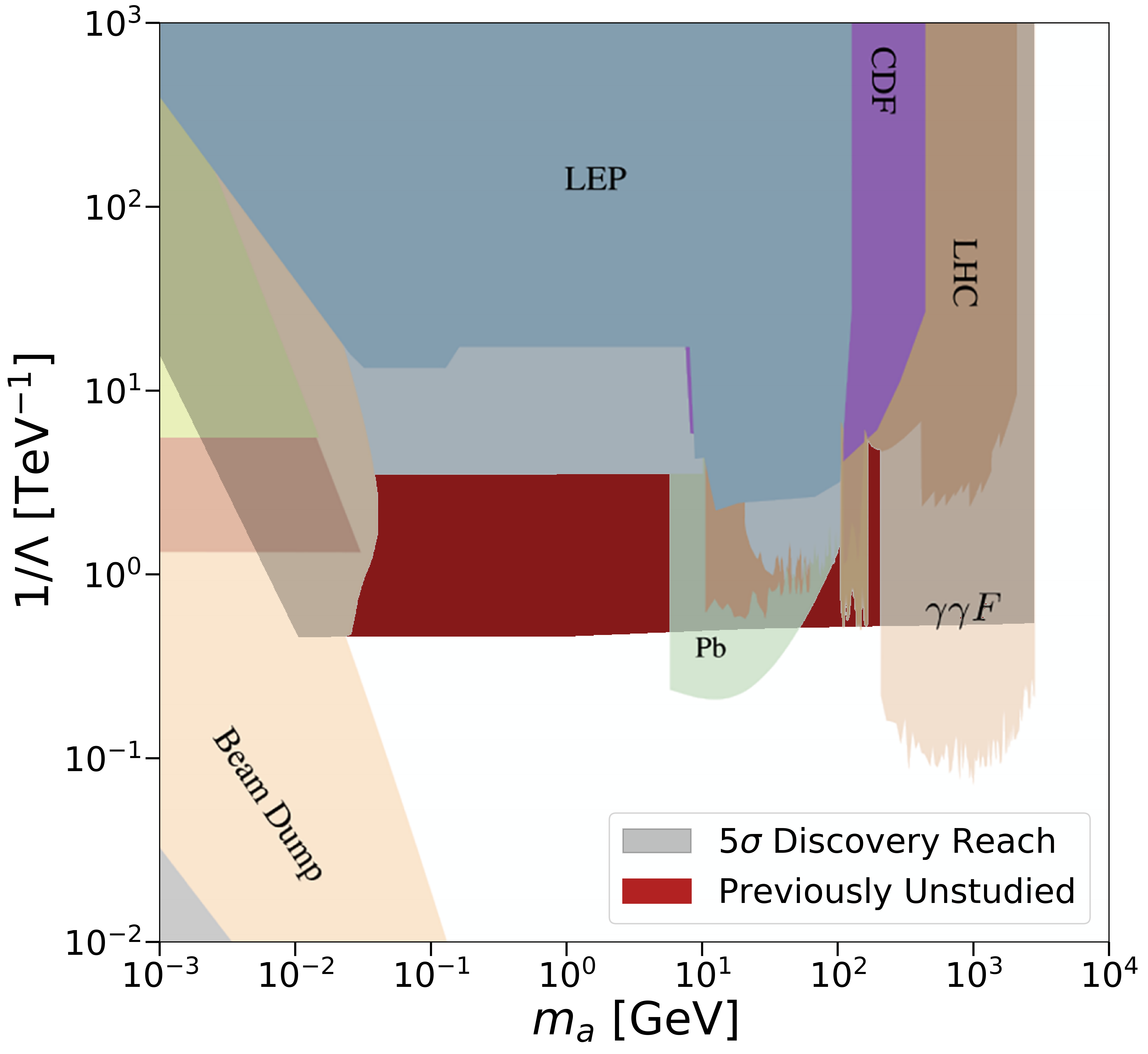}
    \caption{The 5$\sigma$ discovery reach at 3000 fb$^{-1}$ obtained using our search methodology is depicted on ALP parameter space, with an emphasis placed on the subset that has previously not been experimentally probed. The other constraints shown are taken from Figure 4 of \cite{Bauer:2018uxu}.}
    \label{fig:compare}
\end{figure}

Figure~\ref{fig:compare} shows the comparison of our 5$\sigma$ discovery reach at 3000 fb$^{-1}$ to existing constraints on ALP parameter space (grey). 
The constraints shown in Fig.~\ref{fig:compare} are taken from Fig. 4 of \cite{Bauer:2018uxu} and correspond to LEP (light blue and blue), CDF (purple), the LHC (associated production and $Z$ decays (orange), photon fusion (light orange), and  heavy-ion collisions (green)). The results from previous collider searches show a gap in sensitivity in the ALP mass range $10 \text{ MeV} \lesssim m_a \lesssim 100$ GeV, which is primarily due to: $(i)$ low resonant ALP production cross sections at TeV scale values of $\Lambda$; and $(ii)$ the low-$p_{T}$ photon kinematics arising from low mass ALP decays in the traditional searches without a boosted topology, which suffer from large SM backgrounds. It is clear that the proposed methodology using a boosted VBF topology can probe regions of parameter space that are currently unconstrained by other searches (red).

\section{Acknowledgements}

We thank the constant and enduring financial support received for this project from the faculty of science at Universidad de los Andes (Bogot\'a, Colombia), the administrative department of science, technology and innovation of Colombia (COLCIENCIAS), the Physics \& Astronomy department at Vanderbilt University and the US National Science Foundation. This work is supported in part by NSF Award PHY-1806612. KS is supported by DOE Grant DE-SC0009956.


\newpage

\begin{thebibliography}{1}

\bibitem{Weinberg:1977ma}
S.~Weinberg,
``A New Light Boson?,''
Phys. Rev. Lett. \textbf{40}, 223-226 (1978)
doi:10.1103/PhysRevLett.40.223

\bibitem{Wilczek:1977pj}
F.~Wilczek,
``Problem of Strong  $P$  and  $T$  Invariance in the Presence of Instantons,''
Phys. Rev. Lett. \textbf{40}, 279-282 (1978)
doi:10.1103/PhysRevLett.40.279

\bibitem{Peccei:1977hh}
R.~D.~Peccei and H.~R.~Quinn,
``CP Conservation in the Presence of Instantons,''
Phys. Rev. Lett. \textbf{38}, 1440-1443 (1977)
doi:10.1103/PhysRevLett.38.1440

\bibitem{Arvanitaki:2009fg}
A.~Arvanitaki, S.~Dimopoulos, S.~Dubovsky, N.~Kaloper and J.~March-Russell,
``String Axiverse,''
Phys. Rev. D \textbf{81}, 123530 (2010)
doi:10.1103/PhysRevD.81.123530
[arXiv:0905.4720 [hep-th]].

\bibitem{Cicoli:2012sz}
M.~Cicoli, M.~Goodsell and A.~Ringwald,
``The type IIB string axiverse and its low-energy phenomenology,''
JHEP \textbf{10}, 146 (2012)
doi:10.1007/JHEP10(2012)146
[arXiv:1206.0819 [hep-th]].



\bibitem{Sikivie:1983ip}
P.~Sikivie,
``Experimental Tests of the Invisible Axion,''
Phys. Rev. Lett. \textbf{51}, 1415-1417 (1983)
doi:10.1103/PhysRevLett.51.1415


\bibitem{Sikivie:1985yu}
P.~Sikivie,
``Detection Rates for 'Invisible' Axion Searches,''
Phys. Rev. D \textbf{32}, 2988 (1985)
doi:10.1103/PhysRevD.36.974

\bibitem{Anastassopoulos:2017ftl}
V.~Anastassopoulos \textit{et al.} [CAST],
``New CAST Limit on the Axion-Photon Interaction,''
Nature Phys. \textbf{13}, 584-590 (2017)
doi:10.1038/nphys4109
[arXiv:1705.02290 [hep-ex]].

\bibitem{Irastorza:2013dav}
I.~Irastorza \textit{et al.} [IAXO],
``The International Axion Observatory IAXO. Letter of Intent to the CERN SPS committee,''
CERN-SPSC-2013-022.

\bibitem{Spector:2019ooq}
A.~Spector [ALPS],
[arXiv:1906.09011 [physics.ins-det]].

\bibitem{Graham:2015ouw}
P.~W.~Graham, I.~G.~Irastorza, S.~K.~Lamoreaux, A.~Lindner and K.~A.~van Bibber,
``Experimental Searches for the Axion and Axion-Like Particles,''
Ann. Rev. Nucl. Part. Sci. \textbf{65}, 485-514 (2015)
doi:10.1146/annurev-nucl-102014-022120
[arXiv:1602.00039 [hep-ex]].

\bibitem{Fortin:2021sst}
J.~F.~Fortin, H.~K.~Guo, S.~P.~Harris, E.~Sheridan and K.~Sinha,
[arXiv:2101.05302 [hep-ph]].

\bibitem{Harris:2020qim}
S.~P.~Harris, J.~F.~Fortin, K.~Sinha and M.~G.~Alford,
``Axions in neutron star mergers,''
JCAP \textbf{07}, 023 (2020)
doi:10.1088/1475-7516/2020/07/023
[arXiv:2003.09768 [hep-ph]].

\bibitem{Lloyd:2020vzs}
S.~J.~Lloyd, P.~M.~Chadwick, A.~M.~Brown, H.~k.~Guo and K.~Sinha,
``Axion Constraints from Quiescent Soft Gamma-ray Emission from Magnetars,''
[arXiv:2001.10849 [astro-ph.HE]].

\bibitem{Fortin:2018ehg}
J.~F.~Fortin and K.~Sinha,
``Constraining Axion-Like-Particles with Hard X-ray Emission from Magnetars,''
JHEP \textbf{06}, 048 (2018)
doi:10.1007/JHEP06(2018)048
[arXiv:1804.01992 [hep-ph]].

\bibitem{Fortin:2018aom}
J.~F.~Fortin and K.~Sinha,
``X-Ray Polarization Signals from Magnetars with Axion-Like-Particles,''
JHEP \textbf{01}, 163 (2019)
doi:10.1007/JHEP01(2019)163
[arXiv:1807.10773 [hep-ph]].


\bibitem{Buschmann:2019pfp}
M.~Buschmann, R.~T.~Co, C.~Dessert and B.~R.~Safdi,
``X-ray Search for Axions from Nearby Isolated Neutron Stars,''
[arXiv:1910.04164 [hep-ph]].



\bibitem{Dobrich:2015jyk}
B.~D\"obrich, J.~Jaeckel, F.~Kahlhoefer, A.~Ringwald and K.~Schmidt-Hoberg,
JHEP \textbf{02}, 018 (2016)
doi:10.1007/JHEP02(2016)018
[arXiv:1512.03069 [hep-ph]].

\bibitem{Feng:2018pew}
J.~L.~Feng, I.~Galon, F.~Kling and S.~Trojanowski,
``Axionlike particles at FASER: The LHC as a photon beam dump,''
Phys. Rev. D \textbf{98}, no.5, 055021 (2018)
doi:10.1103/PhysRevD.98.055021
[arXiv:1806.02348 [hep-ph]].


\bibitem{Berlin:2018bsc}
A.~Berlin, N.~Blinov, G.~Krnjaic, P.~Schuster and N.~Toro,
``Dark Matter, Millicharges, Axion and Scalar Particles, Gauge Bosons, and Other New Physics with LDMX,''
Phys. Rev. D \textbf{99}, no.7, 075001 (2019)
doi:10.1103/PhysRevD.99.075001
[arXiv:1807.01730 [hep-ph]].

\bibitem{Akesson:2018vlm}
T.~Åkesson \textit{et al.} [LDMX],
``Light Dark Matter eXperiment (LDMX),''
[arXiv:1808.05219 [hep-ex]].

\bibitem{Volpe:2019nzt}
R.~Volpe,
``Search for exotic decays at NA62,''
[arXiv:1910.10429 [hep-ex]].

\bibitem{Berlin:2018pwi}
A.~Berlin, S.~Gori, P.~Schuster and N.~Toro,
``Dark Sectors at the Fermilab SeaQuest Experiment,''
Phys. Rev. D \textbf{98}, no.3, 035011 (2018)
doi:10.1103/PhysRevD.98.035011
[arXiv:1804.00661 [hep-ph]].

\bibitem{Alekhin:2015byh}
S.~Alekhin et. al., Rept. Prog. Phys. \textbf{79}, no.12, 124201 (2016)
doi:10.1088/0034-4885/79/12/124201
[arXiv:1504.04855 [hep-ph]].


\bibitem{Bonivento:2019sri}
W.~M.~Bonivento, D.~Kim and K.~Sinha,
``PASSAT: Particle Accelerator helioScopes for Slim Axion-like-particle deTection,''
Eur. Phys. J. C \textbf{80}, no.2, 164 (2020)
doi:10.1140/epjc/s10052-020-7719-y
[arXiv:1909.03071 [hep-ph]].

\bibitem{Dev:2021ofc}
P.~S.~B.~Dev, D.~Kim, K.~Sinha and Y.~Zhang,
[arXiv:2101.08781 [hep-ph]].

\bibitem{Dent:2019ueq}
J.~B.~Dent, B.~Dutta, D.~Kim, S.~Liao, R.~Mahapatra, K.~Sinha and A.~Thompson,
``New Directions for Axion Searches via Scattering at Reactor Neutrino Experiments,''
Phys. Rev. Lett. \textbf{124}, no.21, 211804 (2020)
doi:10.1103/PhysRevLett.124.211804
[arXiv:1912.05733 [hep-ph]].

\bibitem{Kelly:2020dda}
K.~J.~Kelly, S.~Kumar and Z.~Liu,
[arXiv:2011.05995 [hep-ph]].


\bibitem{Jaeckel:2015jla}
J.~Jaeckel and M.~Spannowsky,
``Probing MeV to 90 GeV axion-like particles with LEP and LHC,''
Phys. Lett. B \textbf{753}, 482-487 (2016)
doi:10.1016/j.physletb.2015.12.037
[arXiv:1509.00476 [hep-ph]].


\bibitem{Brivio:2017ije}
I.~Brivio, M.~B.~Gavela, L.~Merlo, K.~Mimasu, J.~M.~No, R.~del Rey and V.~Sanz,
``ALPs Effective Field Theory and Collider Signatures,''
Eur. Phys. J. C \textbf{77}, no.8, 572 (2017)
doi:10.1140/epjc/s10052-017-5111-3
[arXiv:1701.05379 [hep-ph]].

\bibitem{Bauer:2018uxu}
M.~Bauer, M.~Heiles, M.~Neubert and A.~Thamm,
``Axion-Like Particles at Future Colliders,''
Eur. Phys. J. C \textbf{79}, no.1, 74 (2019)
doi:10.1140/epjc/s10052-019-6587-9
[arXiv:1808.10323 [hep-ph]].

\bibitem{Alves:2016koo}
A.~Alves, A.~G.~Dias and K.~Sinha,
``Diphotons at the $Z$-pole in Models of the 750 GeV Resonance Decaying to Axion-Like Particles,''
JHEP \textbf{08}, 060 (2016)
doi:10.1007/JHEP08(2016)060
[arXiv:1606.06375 [hep-ph]].


\bibitem{Jaeckel:2012yz}
J.~Jaeckel, M.~Jankowiak and M.~Spannowsky,
``LHC probes the hidden sector,''
Phys. Dark Univ. \textbf{2}, 111-117 (2013)
doi:10.1016/j.dark.2013.06.001
[arXiv:1212.3620 [hep-ph]].


\bibitem{vbfatlas1}
ATLAS Collaboration, ATLAS-CONF-2012-091 (2012).


\bibitem{Aad:2012tfa}
G.~Aad \textit{et al.} [ATLAS],
``Observation of a new particle in the search for the Standard Model Higgs boson with the ATLAS detector at the LHC,''
Phys. Lett. B \textbf{716}, 1-29 (2012)
doi:10.1016/j.physletb.2012.08.020
[arXiv:1207.7214 [hep-ex]].

\bibitem{Aad:2012tba}
G.~Aad \textit{et al.} [ATLAS],
``Measurement of isolated-photon pair production in $pp$ collisions at $\sqrt{s}=7$ TeV with the ATLAS detector,''
JHEP \textbf{01}, 086 (2013)
doi:10.1007/JHEP01(2013)086
[arXiv:1211.1913 [hep-ex]].

\bibitem{Florez:2019tqr}
A.~Flórez, A.~Gurrola, W.~Johns, J.~Maruri, P.~Sheldon, K.~Sinha and S.~R.~Starko,
``Anapole Dark Matter via Vector Boson Fusion Processes at the LHC,''
Phys. Rev. D \textbf{100}, no.1, 016017 (2019)
doi:10.1103/PhysRevD.100.016017
[arXiv:1902.01488 [hep-ph]].

\bibitem{DMmodels2}
 A.~Delannoy {\it et al.},
  ``Probing Dark Matter at the LHC using Vector Boson Fusion Processes,''
  Phys.\ Rev.\ Lett.\  {\bf 111}, 061801 (2013)
  doi:10.1103/PhysRevLett.111.061801
  [arXiv:1304.7779 [hep-ph]].
  
  \bibitem{CMSVBFDM} 
  V.~Khachatryan {\it et al.} [CMS Collaboration],
  ``Search for dark matter and supersymmetry with a compressed mass spectrum in the vector boson fusion topology in proton-proton collisions at $\sqrt{s}=8$ TeV,''
  Phys.\ Rev.\ Lett.\  {\bf 118}, 021802 (2017)
  10.1103/PhysRevLett.118.021802
  [arXiv:1605.09305v2 [hep-ex]].
  
  \bibitem{VBF1}
 B.~Dutta, A.~Gurrola, W.~Johns, T.~Kamon, P.~Sheldon and K.~Sinha,
  ``Vector Boson Fusion Processes as a Probe of Supersymmetric Electroweak Sectors at the LHC,''
  Phys.\ Rev.\ D {\bf 87}, no. 3, 035029 (2013)
  doi:10.1103/PhysRevD.87.035029
  [arXiv:1210.0964 [hep-ph]].

\bibitem{VBFSlepton}
 B.~Dutta, T.~Ghosh, A.~Gurrola, W.~Johns, T.~Kamon, P.~Sheldon, K.~Sinha and S.~Wu,
  ``Probing Compressed Sleptons at the LHC using Vector Boson Fusion Processes,''
  Phys.\ Rev.\ D  {\bf 91}, 055025 (2015)
  10.1103/PhysRevD.91.055025
  [arXiv:1411.6043 [hep-ph]].

\bibitem{VBFStop}
 B.~Dutta, W.~Flanagan, A.~Gurrola, W.~Johns, T.~Kamon, P.~Sheldon, K.~Sinha, K.~Wang and S.~Wu,
  ``Probing Compressed Top Squarks at the LHC at 14 TeV,''
  Phys.\ Rev.\ D  {\bf 90}, 095022 (2014)
  10.1103/PhysRevD.90.095022
  [arXiv:1312.1348 [hep-ph]].

\bibitem{VBFSbottom}
 B.~Dutta {\it et al.},
  ``Probing Compressed Bottom Squarks with Boosted Jets and Shape Analysis,''
  Phys.\ Rev.\ D  {\bf 92}, 095009 (2015)
  10.1103/PhysRevD.92.095009
  [arXiv:1507.01001 [hep-ph]].
  
  \bibitem{VBF2}
V.~Khachatryan {\it et al.} [CMS Collaboration],
  ``Search for supersymmetry in the vector-boson fusion topology in proton-proton collisions at $ \sqrt{s}=8 $ TeV,''
  JHEP {\bf 1511}, 189 (2015)
  doi:10.1007/JHEP11(2015)189
  [arXiv:1508.07628 [hep-ex]].
  
 \bibitem{Sirunyan:2019zfq} 
  A.~M.~Sirunyan {\it et al.} [CMS Collaboration],
  ``Search for supersymmetry with a compressed mass spectrum in the vector boson fusion topology with 1-lepton and 0-lepton final states in proton-proton collisions at $\sqrt{s}=$ 13 TeV,''
  JHEP {\bf 1908}, 150 (2019)
  doi:10.1007/JHEP08(2019)150
  [arXiv:1905.13059 [hep-ex]].
  
\bibitem{ConnectingPPandCosmology}
 C.~\'Avila, A.~Andr\'es, A.~Gurrola, D.~Julson and S.~Starko,
  ``Connecting Particle Physics and Cosmology: Measuring the Dark Matter Relic Density in Compressed Supersymmetry at the LHC,''
  Physics of the Dark Universe  {\bf 27}, 100430 (2020), 
  [arXiv:1801.03966 [hep-ph]].

\bibitem{VBFZprime}
 A.~Florez, A.~Gurrola, W.~Johns, Y.~Oh, P.~Sheldon, D.~Teague, and T.~Weiler,
  ``Searching for New Heavy Neutral Gauge Bosons using Vector Boson Fusion Processes at the LHC,''
  Phys.\ Lett.\ B\  {\bf 767}, 126-132 (2017)
  doi:10.1016/j.physletb.2017.01.062
  [arXiv:1609.09765v2 [hep-ph]].
  
\bibitem{VBFHN}
 A.~Florez, A.~Gurrola, K.~Gui, C.~Patino, and D.~Restrepo,
  ``Expanding the Reach of Heavy Neutrino Searches at the LHC,''
  Phys.\ Lett.\ B\  {\bf 778}, 94-100 (2018)
  doi:10.1016/j.physletb.2018.01.009
  [arXiv:1708.03007v1 [hep-ph]].
      
\bibitem{Florez:2018ojp}
A.~Flórez, Y.~Guo, A.~Gurrola, W.~Johns, O.~Ray, P.~Sheldon and S.~Starko,
``Probing heavy spin-2 bosons with $\gamma\gamma$ final states from vector boson fusion processes at the LHC,''
Phys. Rev. D \textbf{99}, no.3, 035034 (2019)
doi:10.1103/PhysRevD.99.035034
[arXiv:1812.06824 [hep-ph]].


\bibitem{Alloul:2013bka}
A.~Alloul, N.~D.~Christensen, C.~Degrande, C.~Duhr and B.~Fuks,
``FeynRules  2.0 - A complete toolbox for tree-level phenomenology,''
Comput. Phys. Commun. \textbf{185}, 2250-2300 (2014)
doi:10.1016/j.cpc.2014.04.012
[arXiv:1310.1921 [hep-ph]].

\bibitem{axionUFO}
https://feynrules.irmp.ucl.ac.be/wiki/ALPsEFT.

\bibitem{MADGRAPH} 
  J.~Alwall {\it et al.},
  ``The automated computation of tree-level and next-to-leading order differential cross sections, and their matching to parton shower simulations,''
  JHEP {\bf 1407}, 079 (2014)
  doi:10.1007/JHEP07(2014)079
  [arXiv:1405.0301 [hep-ph]].
  
 \bibitem{Sjostrand:2014zea}
T.~Sjöstrand, S.~Ask, J.~R.~Christiansen, R.~Corke, N.~Desai, P.~Ilten, S.~Mrenna, S.~Prestel, C.~O.~Rasmussen and P.~Z.~Skands,
``An Introduction to PYTHIA 8.2,''
Comput. Phys. Commun. \textbf{191}, 159-177 (2015)
doi:10.1016/j.cpc.2015.01.024
[arXiv:1410.3012 [hep-ph]].

\bibitem{deFavereau:2013fsa} 
  J.~de Favereau {\it et al.} [DELPHES 3 Collaboration],
  ``DELPHES 3, A modular framework for fast simulation of a generic collider experiment,''
  JHEP {\bf 1402}, 057 (2014)
  doi:10.1007/JHEP02(2014)057

\bibitem{Gavela:2019}
M. B. Gavela \textit{et al.},
``Non-Resonant Searches for Axion-Like Particles at the LHC,''
Phys. Rev. Lett. \textbf{124}, no.5, 051802 (2020)
doi:10.1103/PhysRevLett.124.051802
[arXiv:1905.12953 [hep-ph]].

\bibitem{MLM} 
  J.~Alwall {\it et al.},
  ``Comparative study of various algorithms for the merging of parton showers and matrix elements in hadronic collisions,''
  Eur. Phys. J. C  53, 473 (2008)
  doi:10.1140/epjc/s10052-007-0490-5, arXiv:0706.2569.


 \bibitem{ROOTFit}
L. Moneta, K. Belasco, K. S. Cranmer, S. Kreiss, A. Lazzaro,
et. al., The RooStats Project, PoS ACAT2010 (2010) 057, [1009.1003]

\bibitem{Khachatryan:2016hje}
V.~Khachatryan \textit{et al.} [CMS],
``Search for Resonant Production of High-Mass Photon Pairs in Proton-Proton Collisions at $\sqrt s$ =8 and 13 TeV,''
Phys. Rev. Lett. \textbf{117}, no.5, 051802 (2016)
doi:10.1103/PhysRevLett.117.051802
[arXiv:1606.04093 [hep-ex]].

\bibitem{Aad:2014cka}
G.~Aad \textit{et al.} [ATLAS],
``Search for high-mass dilepton resonances in pp collisions at $\sqrt{s}=8$  TeV with the ATLAS detector,''
Phys. Rev. D \textbf{90}, no.5, 052005 (2014)
doi:10.1103/PhysRevD.90.052005
[arXiv:1405.4123 [hep-ex]].

\bibitem{Butterworth:2015oua} 
  J.~Butterworth {\it et al.},
  ``PDF4LHC recommendations for LHC Run II,''
  J.\ Phys.\ G {\bf 43}, 023001 (2016)
  doi:10.1088/0954-3899/43/2/023001
  [arXiv:1510.03865 [hep-ph]].
  



\end{thebibliography}
\end{document}